# Studying the neural representations of uncertainty

Edgar Y WALKER[*], Department of Physiology and Biophysics, Computational Neuroscience Center, University of Washington, Seattle, WA

Stephan POHL[*], Department of Philosophy, New York University

Rachel N DENISON, Department of Psychological & Brain Sciences, Boston University, Boston, MA

David L BARACK, Departments of Neuroscience and Philosophy, University of Pennsylvania, Philadelphia, PA

Jennifer LEE, Center for Neural Science, New York University

Ned BLOCK, Department of Philosophy, New York University

Wei Ji MA[#] Center for Neural Science and Department of Psychology, New York University

Florent MEYNIEL[#, §], Cognitive Neuroimaging Unit, INSERM, CEA, CNRS, Université Paris-Saclay, NeuroSpin center, Gif/Yvette, France, florent.meyniel@cea.fr

[*]: These authors contributed equally; [#]: These authors jointly supervised this work; [§]: corresponding author

# Abstract

The study of the brain's representations of uncertainty is a central topic in neuroscience. Unlike most quantities of which the neural representation is studied, uncertainty is a property of an observer's beliefs about the world, which poses specific methodological challenges. We analyze how the literature on the neural representations of uncertainty addresses those challenges and distinguish between "code-driven" and "correlational" approaches. Code-driven approaches make assumptions about the neural code for representing world states and the associated uncertainty. By contrast, correlational approaches search for relationships between uncertainty and neural activity without constraints on the neural representation of the world state that this uncertainty accompanies. To compare these two approaches, we apply several criteria for neural representations: sensitivity, specificity, invariance, functionality. Our analysis reveals that the two approaches lead to different, but complementary findings, shaping new research questions and guiding future experiments.

# Introduction

Understanding how the brain represents its environment is one of the major goals of neuroscience and psychology. Another major goal is to understand the uncertainty of these representations[1–5]. Taking into account uncertainty in perceptual processing can be crucial when interacting with the world. Imagine that while hiking through the mountains you have to decide whether to attempt to cross a steep slope. Beside your perception of the slope itself, your uncertainty about the slope should also be taken into account. Perhaps you should move closer in order to reduce your uncertainty before you decide to attempt the crossing. A wide range of human behavior takes into account such uncertainty, including decision making[6–9], learning[10–13], perception[3,14–16] including multi-sensory fusion[17–20], motor control[21,22], and memory[23–26]. Similar observations have been made in non-human animals[27–35].

Many neuroscientists aim to understand how this uncertainty is represented in the brain. Studies of uncertainty often contain claims of the form "in a given brain region, neural activity $r$ represents uncertainty about the latent state $s$". In practice, $r$ can be measured with fMRI, EEG, intracranial recordings of local field potentials, spike trains, amongst others, and $s$ can be the orientation of an object, a reward probability, or some other feature. The goal of this article is to provide a framework for categorizing and evaluating claims about the representation of uncertainty.

# 1. Defining uncertainty

## Uncertainty characterizes the representation of a world state by an observer

Consider some subject who perceives $s$, some feature of interest of the world state. We will understand this situation in terms of a generative model (Fig. 1A and Glossary). The feature $s$ is not directly accessible, and thus called a *latent* feature. The observer receives information about $s$ from the more proximal input state $I$. The brain processes this input to arrive at the neural response $r$ which is a representation of $s$.

In an experimental context, $I$ is the input to the observer that is generated by the feature $s$ in a particular trial. For example, in a visual task, $I$ is the pattern of light that hits the observer's retina, which in practice is considered to be equivalent to the pattern of pixels presented on a screen. As a standard example throughout this article, let $I$ be an image: a grating with orientation $s$ and added random pixel noise (Fig. 1A). Subjects report their estimate of the orientation $s$.

Consider an observer who forms a representation of the world state $s$ through a process that can be described as an inference from $I$ (Fig. 1B, blue arrow). That is, the observer computes values for $s$ given the observed $I$ ($I_{observed}$) and the dependence of $I$ on $s$ assumed in the generative model[36]. However, one and the same input can often be generated from multiple states of the world. That is, the state of the world is underdetermined given the input, leaving the observer uncertain about $s$.

Unlike most quantities that are represented in perception, the uncertainty $u$ about $s$ is not a world state. Rather, the uncertainty $u$ is a property of an observer's belief about the world; $u$ measures the lack of information the observer has about $s$ on the basis of an inference from $I$ [37].

The posterior probability distribution $p(s|I_{observed})$ characterizes the uncertainty about $s$ given $I_{observed}$ in a given trial (Fig. 1C, blue distribution). This distribution describes the probability of different values of $s$ given the particular input $I_{observed\ (Box\ 1)}$. If there were no uncertainty, the value of $s$ would be perfectly determined by some particular input $I_{observed}$, and all the probability mass in the distribution $p(s|I_{observed})$ would be assigned to a single value of $s$. But here, multiple values of $s$ are possible given $I_{observed}$ and thus there is uncertainty about $s$ given $I_{observed}$.

The generative model assumed in an observer's inference need not be the true generative model. The generative process might be too complex for observers to compute an optimal inference[38,39]; they might for instance exclude some variables and simplify the shapes of probability functions[40–42]. Since an observer's uncertainty depends on the generative model assumed in their inference, a major task in the study of representations of uncertainty is to study what generative models are assumed by observers.

Often, an idealized observer is considered who infers $s$ from $I$ based on an optimal inference and the true generative model; the uncertainty such an idealized observer would have about $s$ is called the ideal-observer uncertainty[5,14,43].

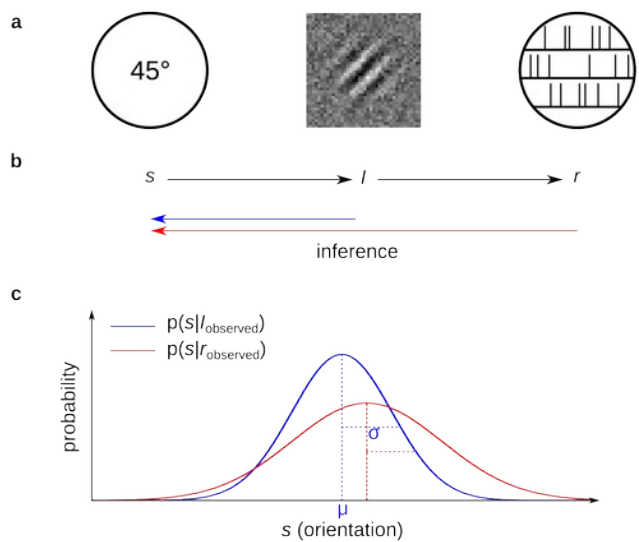

**Figure 1 Uncertainty from a generative model. A**: Example generative model. The world state $s$ is an orientation. The input $I$ is an image of a grating with orientation s and some pixel noise. The neural response $r$ is a train of spikes from neurons in some population. **B**: Black arrows indicate the dependencies in the generative model between the world state $s$, input $I$, and neural response $r$. Red and blue arrows indicate inferences an observer could make by inverting dependencies in the generative model from $I$ and $r$, respectively. **C**: Probability distributions of $s$ given the particular input $I_{observed}$ or the particular neural response $r_{observed}$ in a given trial. σ is a measure of the uncertainty about s given $I_{observed}$, and μ is the expected value of $s$ given $I_{observed}$.

## Origins of uncertainty

Uncertainty about the world state $s$ is present whenever $s$ is under-determined given the inference an observer performs. This under-determination manifests as a many-to-one mapping from the world state to a later state in the generative model assumed by the observer.

One source of under-determination is ambiguity of the input, illustrated for instance by the case of the Necker cube[44]. One and the same two-dimensional image could be interpreted to be the result of different states of the three dimensional world, leaving the observer uncertain between these different states.

Another common source of uncertainty is randomness in the way an earlier state in a generative model generates a later state. One important type of such randomness pertains to the input the observer receives, such as the random pixel noise that corrupts the image of a grating in Fig. 1. An observer is left uncertain when *I* under-determines *s* because of noise.

Another important type of randomness is internal to the observer. The firing behavior of neurons is only partially driven by the signal they receive and partially by further random factors[45]. One and the same neural response *r* can be caused by different input states *I* which in turn depend on different world states *s*. An actual observer does not infer *s* from *I* but from *r*; there is generally more uncertainty about *s* given *r* than given *I* due to the additional uncertainty about *I* given *r* (Fig. 1C, red distribution).

The extent to which randomness increases uncertainty depends on the amount of data available to the observer. For instance, when there is a fixed amount of pixel noise, the orientation task in Fig. 1 is easier for images with more pixels. Similarly, uncertainty about *s* decreases when evidence can be accumulated across multiple sensory inputs *I* and it is large when only little evidence about *s* is observed, such as when *s* changes across time[11,46–49].

### Box 1: Measures of uncertainty

Different formal measures are available to summarize uncertainty *u*. Often, uncertainty is understood as the standard deviation of a random variable under a posterior distribution, especially for the frequently used Gaussian distribution (see Fig. 1C). The larger the standard deviation, i.e. the more spread out the probability distribution, the more uncertainty there is about this variable.

Another useful measure of uncertainty is entropy. The entropy[37] of the posterior distribution $p(s|I_{observed})$ is a measure for how much freedom of choice (hence, uncertainty) there is left about the variable *s*, after one already knows that the variable *I* takes the value $I_{observed}$. The advantage of entropy as a measure of uncertainty is that it applies to probability distributions of any shape (categorical and numeric variables, and with one or more dimensions). However, the fact that entropy ignores ordinality can be a disadvantage: for instance, if two orientations have high probability and all other orientations have the same low probability, entropy (unlike standard deviation) will be the same no matter whether those two orientations are very close or very far apart.

Yet rather than summarizing uncertainty in a single quantity, one might also keep track of it implicitly in terms of the full probability distribution. If one were to represent the state of the world *s* in terms of, for instance, the posterior distribution $p(s|I_{observerd})$, uncertainty about s would be implicit in that representation. This uncertainty can be taken into account implicitly by performing computations on the full probability distribution[92]. Whether observers use full distributions or summaries is an open empirical question[58,108,124].

# 2. Studying representations of uncertainty with correlational and code-driven approaches

Empirical studies on the neural representation of uncertainty differ along various dimensions, such as the recording technique used, the species, the tasks. Here we propose a distinction that reflects a difference in approaches to uncertainty of distinct research communities, one rooted in cognitive neuroscience and the other in theoretical neuroscience. At the methodological level, this distinction is based on whether assumptions about neural coding of world states are used to study uncertainty.

The correlational approach to the representation of uncertainty does not make any (explicit) assumption about a neural code for representing *s*. Instead, researchers use a proxy for the brain's uncertainty about *s* that they derive from the input or behavior (denoted *u*(*I*) and *u*(*b*) respectively). This proxy then guides the search for parts of the brain whose activity *r* co-varies with it. This search is loosely constrained by assumptions about the neural code of *u*. For instance, in a study that measures *r* with fMRI, a relation between *u* and *r* is typically tested for every single voxel in the brain.

The code-driven approach, by contrast, makes strong assumptions about the neural code of the representations of the latent world state *s* and the associated uncertainty *u*. Based on those assumptions, researchers can read out *u* from neural activity *r*. The model is tested by relating the uncertainty derived from the neural activity (denoted *u*(*r*)) with estimates of uncertainty derived from the sensory input (*u*(*I*)) or the behavioral response *b*.

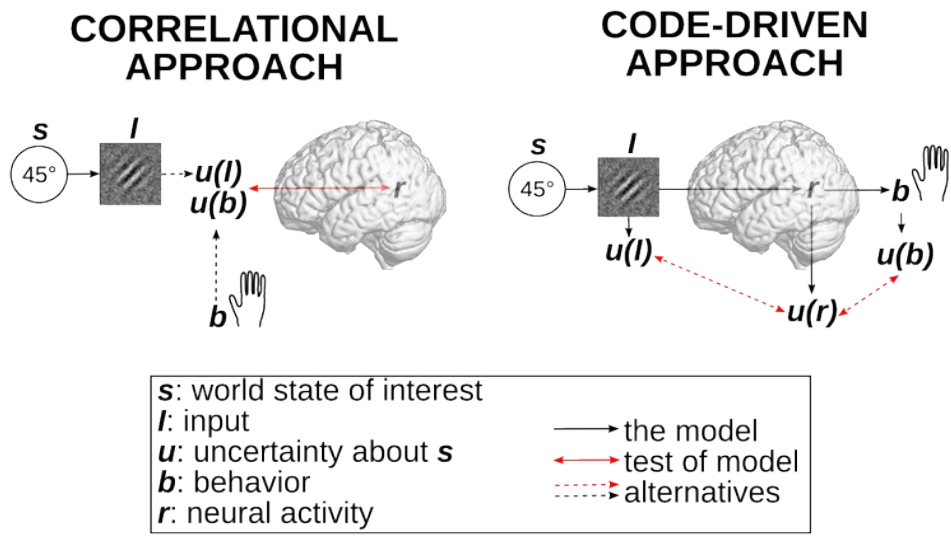

**Figure 2: Comparison of the correlational and code-driven approaches.** In both approaches, the subject is provided with an input *I* (here, an image of a grating) that is informative about a particular world state *s* (here, the orientation of the grating), the subject may provide a response (denoted by the hand), and researchers use some estimates of the brain's uncertainty *u* about *s*. Those estimates are derived from *I* itself (denoted *u*(*I*), based on the modeling of $p(I|s)$ or simple qualitative relationship between *u* and *I*) or *b* (denoted *u*(*b*), based on the subjects' report of *u* or aspects of their choices that should depend on *u*). In the correlational approach, researchers test for a relationship (red arrow) between the proxy *u*(*I*) or *u*(*b*) and neural activity *r*, without making assumptions about the relationship between *s* (or *I*) and *r*. By contrast, the code-driven approach assumes a specific neural code for *s* (through *I*), denoted by the black arrow from *I* to *r*. Based on those assumptions, researchers can read out *u*(*r*), the uncertainty about *s* given the observed *r*, see black arrow from *r* to *u*(*r*). In practice, *u*(*r*) is obtained by inverting a neural likelihood function $\mathcal{L}(s; r_{observed}) = p(r_{observed}|s)$ in a probabilistic population code (PPC) or by measuring the standard deviation of *r* in a sampling-based code (SBC). The validity of the code-based approach and the neural readout of uncertainty is evaluated by testing for some relationship between *u*(*r*) and either *u*(*b*) or *u*(*I*).

## Prototypical examples of the two approaches

We illustrate the correlational approach with an fMRI study from Vilares and colleagues[50]. In this study, subjects were presented on each trial with an input $I$ consisting of a dot-cloud sampled from a Gaussian distribution whose mean location was the latent state $s$ that subjects had to report. The experimenters used two dispersion levels for the dot-cloud (i.e. variance of the Gaussian distribution), small or large, to manipulate the subject's uncertainty about $s$. The authors asked where the input-related uncertainty $u(I)$ (using the dispersion of the dot-cloud as a proxy) was represented in the brain, and stressed the absence of strong hypothesis: "it remains unknown whether uncertainty is represented along the sensorimotor pathway or within specialized brain areas outside this pathway". To detect and localize a representation of uncertainty, they regressed $u(I)$ against the fMRI signal in each voxel throughout the brain. They found that fMRI activity positively correlated with $u(I)$ in the early visual cortex.

As illustration of the code-driven approach, consider a study from Geurts and colleagues[51] who presented subjects with an input $I$ consisting of an oriented grating (the orientation is the latent state $s$) with low contrast. Subjects reported both the orientation of the grating and their uncertainty. The authors made very specific assumptions about the neural representation of $s$ (the orientation) in the early visual cortex: "The model assumes that, across trials, voxel activity follows a multivariate normal distribution around the voxel's tuning curve for orientation." When fitted to the fMRI data, this model characterizes the probability of an activity pattern $r$ given the orientation $s$. Using probabilistic inference, this model can be inverted to estimate the probability distribution of $s$ given the observed $r$ (as illustrated in Fig 1), and the uncertainty about $s$ inferred from $r$, $u(r)$, can be measured as the standard deviation of this distribution. To test whether it is indeed a neural representation of uncertainty about $s$, the authors regressed across trials $u(r)$ against $u(b)$, the uncertainty reported by subjects, and found a significant positive effect.

## Estimates of uncertainty derived from the input and behavior

The above prototypical examples illustrate that the two approaches use some estimate $u(I)$ and $u(b)$ of uncertainty derived from the input or behavior and provide specific examples of such estimates; below we provide a broader range of examples. In general, the same estimates of $u(I)$ and $u(b)$ are available for either approach. What distinguishes the two approaches is the way those estimates are used: as a proxy for the brain's uncertainty to search for neural representations of this uncertainty in the correlational approach, and as a check of the neural readout of uncertainty in the code-driven approach.

Different aspects of behavior are used to derive estimates of uncertainty. In some studies, $u(b)$ is the uncertainty reported by human subjects (e.g. ratings or confidence judgments[46,47,51–60]). In other studies, researchers infer $u(b)$ by assuming that uncertainty regulates some specific aspects of subject's behavior[58,61], such as how fast to respond[62,63], how long to wait for a reward[30,64,65], whether to opt out of a bet[31,32,34,66–68], the variability of behavioral reports[57,69] or the relative weight between prior information and the current input $I$[50,70].

Different methods also exist to estimate uncertainty from the input. Some researchers use ideal-observer models (see section 1), which are useful to quantify uncertainty across a wide variety of task structures, notably when the relation between $u$ and $I$ is complex, e.g. in sequential learning[7,46,47,71–73]. Other models used to estimate $u(I)$ go beyond the task-based generative model and incorporate assumptions about the decomposition of $I$ into specific features by sensory systems[74–76]. Another method eschews the use of generative models of $I$ by relying on simple qualitative relationships that exist between $u$ and specific aspects of $I$ (e.g. pixel noise or contrast in the oriented grating example, or the fact that humans are more uncertain about oblique than cardinal orientations); these are used as crude estimates of uncertainty[30,31,69,77].

## Assumptions about the neural code

The correlational approach seeks to relate the uncertainty proxy $u(l)$ or $u(b)$, to $r$ (see Fig. 2). In practice, researchers test for this relationship by different means, such as correlation, multiple linear regressions or multivariate pattern decoding (e.g. with support-vector machines, Box 2). Each method implicitly makes assumptions about the neural code of $u$ (e.g. linearity in the case of correlation) but the choice of a method tends to be motivated more by convenience (the use of standard tools that capture simple statistical relationships between $u$ and $r$) than strong hypotheses about the code. Depending on the method, the strength of this relationship is measured as a correlation coefficient, the significance of regression weights[78], the cross-validated decoding accuracy, or the fraction of explained variance[79,80].

By contrast, the code-driven approach makes assumptions (in the form of a "neural" generative model) about how $r$ represents $s$. We present two families of such models[81]: those that posit a neural code for $s$ in which $r$ encodes a likelihood function $\mathcal{L}(s;r)=p(r|s)$ (broadly referred to as a Probabilistic Population Code) and those that posit a neural code for $s$ in which $r$ corresponds to samples from a posterior distribution over $s$ given the observed $I$, $p(s|I_{observed})$ (Sampling-Based Code). Note that those two approaches are not necessarily contradictory[82]. Other models relevant for the study of uncertainty exist[3,83–87], but have so far received less attention from experimenters.

In Probabilistic Population Codes, researchers formalize the uncertainty $u$ conveyed by the neural activity observed on a particular trial, $r_{observed}$, as the posterior distribution $p(s|r_{observed})$. This posterior is derived using Bayes' rule from a neural likelihood function $\mathcal{L}(s;r_{observed})$ (and a prior over $s$ but this aspect is typically obviated by assuming non-informative priors). The construction of $\mathcal{L}(s;r)$ can be more or less data-driven. The dominant approach in the literature is strongly theory-driven[88,89]. An influential example in the sensory domain posits that neurons have a stereotyped mean response to the input (known as their tuning curve) and some variability corresponding to the exponential family of distributions (e.g. Poisson distributions). Together with a few other assumptions, the log of $\mathcal{L}(s;r)$ becomes linear with respect to $r$ and uncertainty about $s$ is proportional to the average neural activity on a given trial[90–92]. By contrast, more data-driven approaches require fewer assumptions and estimate $\mathcal{L}(s;r)$ from the data itself. With the advent of large datasets and machine learning tools, even arbitrary shapes of $\mathcal{L}(s;r)$ can be estimated[35]. With a smaller amount of data, further constraints are needed about the shape of $\mathcal{L}(s;r)$, e.g. assumptions about specific covariance matrices or noise distributions[51,57,69,70].

In Sampling-Based Codes, the neural activity is assumed to represent $s$ in terms of samples stochastically drawn from the posterior distribution $p(s|I_{observed})$[2,4,76,93]. Biologically plausible neural network models have been proposed for such a sampling process[94]. Under such a code, $u$ is reflected in the spread of the distribution of $r$ (e.g. the standard deviation) across time or across neurons.

To summarize, studies within the correlational and code-driven approaches to studying uncertainty differ across many dimensions (the recording techniques; whether mathematical models, and $I$ or $b$, are used to estimate $u$). The key difference is whether assumptions about the neural code of $s$ are used to relate $r$ to the uncertainty about $s$.

## Box 2: Relation to encoding and decoding approaches

Encoding and decoding are widely used notions in neuroscience[79,81,88,125–129]. The encoding approach models *r* as a function of some task-related quantity *x* (typically *s* or *I*, or *u*(*I*) or *u*(*b*) in the context of uncertainty); the decoding approach models *x* as a function of *r*. The relationship between encoding/decoding and correlational/code-driven is multifaceted, notably because in practice different implementations of encoding/decoding exist in the domains of data analysis (machine learning) and theoretical neuroscience.

In theoretical neuroscience, encoding/decoding models are expressed in terms of conditional probabilities[88,129,130]. An encoding model corresponds to $p(r|s)$ (and thus the neural likelihood function $\mathcal{L}(s;r)$), a decoding model corresponds to $p(s|r)$ (and thus potentially captures the brain's uncertainty about *s*). It is possible to obtain $p(s|r)$ from $p(r|s)$ together with a prior probability $p(s)$ using Bayes' rule, which indicates that encoding/decoding models are related but not equivalent since a prior is also involved. The essence of Probabilistic Population Codes is to model $\mathcal{L}(s;r)$ (encoding) and use it to obtain $p(s|r)$ (decoding), thus both encoding and decoding can be related to a code-driven approach to uncertainty.

By contrast, encoding and decoding in the data analysis domains[79,128] and machine learning applied to neuroscience[126,127,131] typically do not involve conditional probabilities between *s* and *r*. In this domain an encoding model typically corresponds to first assuming a deterministic mapping (often highly non-linear) from *s* (or *I*) to a list of latent features, and then testing for a relation to *r* by means of a (multiple) linear regression of the latent features onto *r* (*linearizing encoding models*[128]). Some studies that follow the correlational approach to uncertainty use this encoding method because they treat the uncertainty $u(I)$ as a latent feature of the input *I*, and regress $u(I)$ onto *r*. The same is true of sampling-based code studies except that they consider the variability of *r* rather than *r* itself. Regarding decoding models, they typically correspond to classifiers (e.g. linear discriminant analysis or support vector machine) that are trained to obtain *s* (or *I*) from *r*. This method is also used by some studies that follow the correlational approach, with the twist that the classifier is trained to obtain $u(I)$ or $u(b)$ instead of *s* itself.

Some applications of encoding/decoding are hybrid and use both a decomposition of *s* (or *I*) into some latent features that are regressed onto *r*, and conditional probabilities to model $p(r|s)$ using the residuals of this regression. Some researchers in the code-driven approach used such a method to parametrize a probabilistic encoding model $p(r|s)$ and then decode the uncertainty about *s* given $r$[51,57,69,70].

# 3. General criteria for the evaluation of claims about representations of uncertainty

We now turn to the evaluation of claims about the neural representation of uncertainty that can be found in either the correlational or code-driven approach. We propose to do so by applying criteria that are generally used in neuroscience to support claims about representations: sensitivity, specificity, invariance and functionality (Box 3).

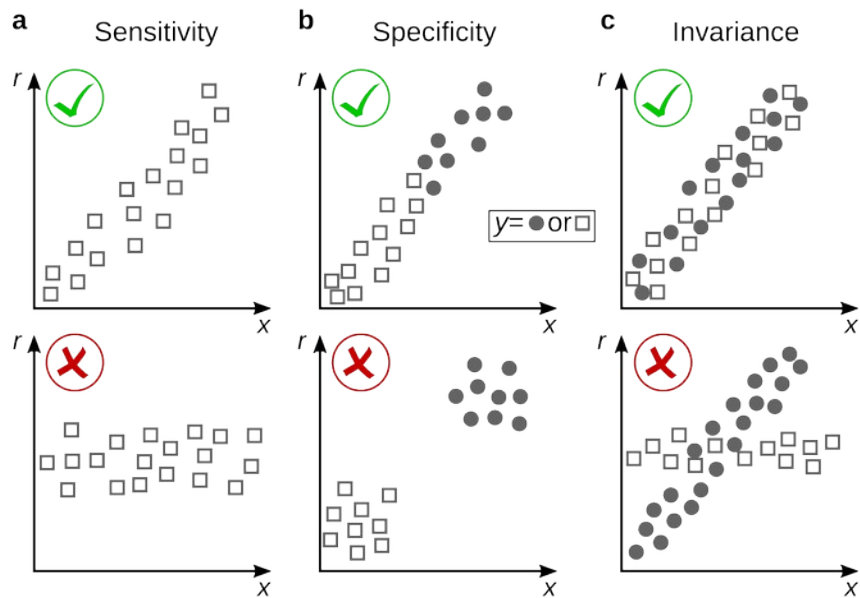

**Figure 3: Empirical criteria for neural representation.** We consider three criteria (sensitivity, specificity and invariance) that are used in neuroscience to test for whether neural activity $r$ represents a feature of interest $x$ (like uncertainty, derived either from some input to the subject, $u(I)$, or from their behavior, $u(b)$), either by itself or in comparison to another feature $y$. The top row shows examples that pass a given criteria and the bottom row examples that fail. Note that in contrast to the correlational approach, the code-driven approach is interested in the uncertainty derived from $r$, $u(r)$, rather than $r$ itself; in that case, $u(r)$ replaces $r$ in the above graphs. Also note that the other feature $y$ is categorical in this figure, but it could also be continuous.

## Box 3: General criteria for neural representations

Below we list criteria generally used in neuroscience to establish claims about representations (although they do not always appear under the labels we propose).

**Sensitivity:** $r$ is sensitive to a feature $x$ if changes in $r$ are related to changes in $x$. For instance, a neuron is sensitive to the orientation of a bar if different activity patterns are recorded when different orientations are presented[132].

**Specificity:** $r$ represents $x$ specifically with respect to another feature $y$ if changes in $r$ are related to changes in $x$ even when controlling for $y$. This criterion enables researchers to test that $r$ is related to $x$ indeed, and not spuriously so because of another feature $y$ that is related to $x$ (in that case, $y$ is termed a confounding variable)[133]. For instance, uncertainty about orientation depends on the image contrast: a neural representation of the uncertainty about orientation therefore ought to be sensitive to contrast. However, to be a representation of uncertainty *per se* rather than contrast, $r$ should reflect uncertainty even when the image contrast is kept fixed.

**Invariance:** The representation of $x$ by $r$ is invariant to $y$ if changes in $r$ are not related to changes in $y$ when controlling for $x$. This criterion enables the researcher to test that $r$ is related to $x$ because $r$ does not change when a feature $y$ unrelated to $x$ changes. For instance, the representation of orientation (our $x$) by V1 neurons (our $r$) is not invariant to position since different $r$ are observed for a given orientation when changing position in the visual field[132]. By contrast, the representation of object identity in the infero-temporal cortex is invariant to the position and orientation of objects[116,134].

**Functionality:** $r$ is functional as a representation of $x$ if it causes a behavioral response $b$, e.g. a report of the perceived value of $x$[135,136]; or, in the case of uncertainty, a decision which weighs sources of information by their respective uncertainty[50]. One can test for functionality with criteria analogous to the ones presented above; yet rather than testing for a dependence of $r$ on $x$, these functionality analogues test for a dependence of $b$ on $r$.

Claims about representations, ultimately, have to be claims about the causal structure of information processing in the brain. Nonetheless, we express the criteria in terms of information theoretic relationships between variables rather than in causal terms because researchers often use correlative (not causal) methods. Previous studies on uncertainty have used correlation[29,51,69], differences between conditions[31,32,50], linear regression[46,47,49,55], or decoding[55,103,107,137,138] to establish representations of uncertainty.

In practice, the criteria are graded rather than all-or-none (sensitivity might for instance be measured in terms of the strength of the correlation between $r$ and $x$). Moreover, while linear relationships such as those illustrated in Fig. 3 are often simplest to understand and most common in experimental studies, many models, especially code-driven models, posit non-linear relationships.

## How do the two approaches compare in terms of testing general criteria?

**Sensitivity**. Testing for sensitivity is the starting point of the correlational approach because it aims to identify whether some $r$ in the brain is sensitive to a proxy $u(I)$ for uncertainty (for $u(b)$, see Functionality below). The higher the sensitivity, such as the strength of correlation or decoding accuracy, the more plausible it is that a given $r$ is a representation of $u$. The use of better proxies for uncertainty also makes tests of sensitivity more convincing. Some studies that follow the correlational approach only address sensitivity, especially when they are among the first of their kind or when uncertainty is not central in the study[48,49,95].

The code-driven approach tests whether a neural readout of uncertainty $u(r)$ is sensitive to $u(I)$. In this case, the vertical axis in Fig. 3 is $u(r)$, not $r$ as in the correlational approach, though in both cases $x=u(I)$. To illustrate, $u(r)$ was shown to be sensitive to aspects of $I$ that impact uncertainty, such as the image contrast[35,76,94], whether orientation is closer to cardinal axes[69], or the presence of a higher-level context[96]. Although either a probabilistic population code or a sampling-based code was used to derive $u(r)$ in those example studies, a prominent difference is that the sensitivity test is more central in studies that use a sampling-based code. In the case of the probabilistic population code, sensitivity sometimes appears as a side point[35] or is even not reported[70].

**Specificity**. In both approaches, researchers correlate (or use more elaborate analyses) $u(I)$ or $u(b)$ to $r$ or $u(r)$. Testing for correlation is vulnerable to the problem of confounding variables: $r$ may not represent $u$ but the aspect of $I$ or $b$ from which $u(I)$ or $u(b)$ has been derived, such as contrast in the orientation task. It is still possible to test for specificity if several features of $I$ or $b$ are related to $u$. In that case, specificity of $r$ to $u$ with respect to each feature in isolation can be tested by holding each feature fixed and testing for the dependence of $r$ on other features of $I$ or $b$[27,74,77]. For example, Dekleva and colleagues[27] manipulated uncertainty about reaching direction through the current trial's cue and the cue history, and Bang and colleagues[77] manipulated uncertainty about the direction of motion by changing the strength of motion evidence and the distance to the category boundary. In both studies, $r$ continued to track $u(I)$ when either feature was kept fixed. Some aspects of $I$ can be artifactually correlated with $u$. For instance, uncertainty about local features in an image is expected to decrease when they are embedded in a higher-level structure; one can test for this effect while controlling for the spectral content of the image, which is often confounded with the presence of high-level structure[96]. Some previous studies include tests for confounding variables such as reaction times[97], attention[97], exploration[7,98], and task difficulty[30,31,64,66,77].

When $u(I)$ is not derived from a simple feature of the input but from a more complex model, like an ideal-observer model, several confounding variables might still undermine the specificity of $r$ to $u$. For instance, in the context of sequential learning, $u$ is often negatively correlated with recently surprising outcomes[46,47,99]. Confounding variables of uncertainty about the present world state also include constructs presented in other studies like the likelihood of a change point[11], expected uncertainty[100], total uncertainty[9,71], outcome uncertainty[46,47,101], and expected reward[102–105].

**Invariance**. Some researchers following the correlational approach have tested for invariance. For instance, Michael and colleagues[106] used a categorization task and inputs with two features: shape and color. The relevant feature used for the categorization task changed across trials and a common neural representation $r$ of the categorization uncertainty was found in both conditions. In the shape condition, $r$ tracked the uncertainty related to shape, not color (and vice versa in the color condition), demonstrating that $r$ coded for uncertainty beyond these low-level features. Using a similar logic, Lebreton and colleagues[56] found a general neural representation of the uncertainty associated with estimating the value of paintings, objects, and prospects. Other researchers have tested invariance with respect to sensory modality[47,64,107].

Invariance is rarely tested in the code-driven approach. Orban et al found that $u(r)$ (in this case, the variability of the membrane potential) was sensitive to some $u(I)$ (the contrast of a grating) and tested for invariance with respect to the orientation of the grating. They reported "mild modulations" by orientation when $u(I)$ was kept fixed [76].

**Functionality**. The correlational approach can use $u(b)$ instead of $u(I)$ as just another proxy for $u$. In that case, functional sensitivity (tested as a relation between $r$ or $u(r)$ and $u(b)$) is not fundamentally different from the sensitivity test presented above (based on $u(I)$). However, $u(I)$ or $u(b)$ are unlikely to be equivalent and it is unclear whether one is a better proxy for the brain's $u$ than the other because the ability of participants to introspect $u$ may be limited[108,109] and additional processes may intervene between $u$ and choices or reports based on $u$[60,110]. Instead of using $u(b)$ in place of $u(I)$ as just another proxy for $u$, more compelling evidence of functionality comes from correlational studies that combine $u(I)$ and $u(b)$. For instance, a neural representation $r$ of $u$ based on $u(I)$ is identified first and then some relation between this $r$ and $b$ is sought. Some studies reported correlations between neural representations of $u(I)$ and the reported uncertainty[46], trial-to-trial variability during reaching[27] and learning[11]. Other studies reported correlations across subjects between neural representations of $u(I)$ and aspects of behavior that should in principle depend on uncertainty, like risk attitude[111], exploration[9,71], and in prior-likelihood combination[50,112].

In the context of the code-driven approach with a probabilistic population code, the functionality criterion often plays a key role. For instance, van Bergen and colleagues[69] showed that $u(r)$, the uncertainty about the orientation of a grating inferred from V1 fMRI activity, correlated with the variability of orientation reports. The same group also found that this $u(r)$ correlated across trials with both the uncertainty reported by subjects[51] and the strength of sequential effects in their perceptual decisions [70]. Walker and colleagues[35] found that the uncertainty read out from V1 in monkeys accounted for their decisions in an uncertainty-based categorization task.

In contrast, functionality is less often tested in studies that assume a sampling-based code. An exception is Haefner and colleagues' 2016 study[113], which reported that the structure of covariance among artificial neurons in a network reflected the uncertainty about the task-relevant orientation during visual categorization in a way that correlated with performance in the task.

Functionality can be coupled with specificity. Such a test assesses whether $r$ or $u(r)$ still correlates with $u(b)$ when all features of the input $I$ are held constant. Uncovering such an effect suggests that it is indeed uncertainty, and not any confounding feature of the input, that is represented and used for behavior. The code-driven approach lends itself to such a test because researchers can read out $u$ from $r$, making it possible to find correlation between $u(r)$ and $u(b)$ when $I$ is fixed[35,69]. The correlational approach does not seem suited for this test when it uses $u(I)$. But, some analyses inspired by this test are still informative; for instance, McGuire and colleagues used $u(I)$ from an ideal observer to identify a neural representation $r$ of $u(I)$; they then regressed $u(I)$ out of $r$ (which is analogous to keeping the effect of $I$ fixed) and showed that $r$ still correlated with some aspect of behavior[11]. Showing that $r$ or $u(r)$ is sensitive to $u(I)$, and to $u(b)$ on top of the effect of $u(I)$, indicates that the way the brain computes uncertainty based on the input deviates from the model assumed to derive $u(I)$, or that other processes (e.g. internal noise, attention[114], or biases) intervene in the computation of uncertainty or its use in behavior.

## How do the two approaches compare in terms of satisfying general criteria?

Supplementary Table 1 reports examples of previous studies on the neural representation of uncertainty. The uncertainty that characterizes representations of the world by an observer is often distinguished from the uncertainty about the outcome of a process[95,101] and from decision confidence[60]. Yet, the corresponding studies often test the same criteria and face similar methodological problems; therefore we include all of them in this table.

Supplementary Table 1 shows how the different approaches (correlational, code-driven) and the different types of uncertainty compare in terms of satisfying the criteria. The criteria apply to both approaches, but with some differences: functionality is currently a strong point of the code-driven approach (when it relies on probabilistic population codes, not on sampling-based codes) in comparison to the correlational approach, whereas invariance, and to a lesser extent specificity, are more often tested in the correlational approach than the code-driven approach.

# 4. Caveats of current approaches and future directions

We now summarize the potential and limitations of the two approaches.

## Comparing correlational and code-driven approaches

**Assumptions about neural codes.** We based our distinction between code-driven and correlational approaches on whether assumptions are made about the neural coding of the world state $s$ and the accompanying uncertainty. This methodological difference has conceptual implications. The code-driven approach studies neural representations in which a neural population $r$ jointly represents both a world state $s$ and uncertainty $u$ about $s$. The correlational approach does not require such joint representations and therefore, it can identify representations of $u$ that are not colocalized with the representation of $s^{50}$. It has been proposed that some brain regions could be specialized in the representation and processing of uncertainty[58,101,108,115]; such brain regions can be identified with the correlational approach, but not with the code-driven approach in its current form.

This restriction to joint representations is likely to explain different findings between the two approaches. The code-driven approach more often identifies representations of $u$ in sensory regions like the early visual cortex, which are well known for representing visual features[35,51,69,70,75,76], whereas the correlational approach often identifies representations of $u$ in regions that are further from sensory input and its representation, closer to the decision or reporting mechanisms, whether from subcortical structures[49,50,97,112], prefrontal cortex[11,46,49,50,56,71,98,106], parietal cortex[11,47,49,71,98,106] or temporal cortex[71,111].

The two approaches also differ in assumptions about the complexity of the neural code of uncertainty. In the correlational approach, codes are usually assumed to be linear (monotonic changes in average activity as a function of uncertainty); representations with such linear codes have been termed explicit representations[116,117]. Studies following the code-driven approach are open to non-linear computations, which are often employed to derive $u$ from $r$, e.g. when reading out the standard deviation of the decoded distribution[51,69,70], or the standard deviation of neural activity[76], or when using artificial neural networks[35].

**Specificity**. Both approaches use external estimates of uncertainty $u(I)$ and $u(b)$ and thus, are susceptible to confounding factors. The code-driven approach has an elegant method to demonstrate functional specificity with respect to the features of the input $I$, namely by testing whether $u(r)$ makes a difference to the behavioral response $b$, even while controlling for $I^{35}$. Yet, concerns about specificity remain even in this case because representations of uncertainty may still be confounded by behavioral features or processes internal to the brain, such as attention.

**Functionality.** Behavioral data are used with substantial heterogeneity in both approaches. In the code-driven approach in particular, it is striking that $u(b)$ is much more prominent than $u(I)$ in studies using probabilistic population codes (PPC), and that the converse is true in those using sampling-based codes. Interestingly, tests based on $u(b)$ have fewer degrees of freedom and thus seem more stringent in the (PPC) code-driven approach than in many studies following the correlational approach. To illustrate, the test is passed in the code-driven approach only if $u(r)$ correlates with $u(b)$, the participant's report of uncertainty[51], whereas it is passed in the correlational approach if $r$ correlates with $u(b)$ in at least one of the many brain regions under investigation.

**Origins of uncertainty**. Externally generated uncertainty $u(I)$ derives from ambiguity or noise in the generation of the sensory input $I$ and can be estimated with an ideal observer model. Internally generated uncertainty depends on neural noise or limitations and errors in information processing; it can only be estimated from behavioral responses $u(b)$ or neural activity $u(r)$; $u(b)$ and $u(r)$ also track external sources of uncertainty. Studies following the correlational approach that focus on $u(I)$ are restricted to externally generated uncertainty. However, by including $u(b)$ in the analysis, correlational studies can also account for internally generated uncertainty. Because they read the uncertainty $u(r)$ directly from the brain state $r$, code-driven models are especially well suited to study internally generated uncertainty. Interestingly, studies using sampling-based codes currently assume that those internal sources of uncertainty (noise) are negligible and that $u(r)$ correspond to the uncertainty optimally computed from $I^{76}$; by contrast, studies using probabilistic population codes stress the importance of internal sources of uncertainty[92,118].

## Setting goals for future research

Given that the code-driven and correlational approaches have different limitations and advantages, they could be used in synergy. One possibility is to leverage our knowledge of early sensory cortices to have a neural readout of uncertainty $u(r)$ about $s$ in a perceptual task using the code-driven approach, and then use $u(r)$ as input to the correlational approach in order to unravel other parts of the brain that could represent this uncertainty without requiring that they represent $s$ itself. Such combined analysis would reconcile the fact that the representation of uncertainty can be colocalized with the representation of the feature $s$ that it characterizes while also being detached from it by downstream computation. Some studies have already started to reduce the gap between code-driven and correlational approaches. The study by Geurts et al.[51] that we used as a prototypical example of the code-driven approach also used the correlational approach and found fMRI correlates of $u(r)$ in the prefrontal cortex.

Understanding how the brain extracts and uses uncertainty can also be achieved by further investigation of the functional aspect of representations. If uncertainty is used only in a given context (e.g. uncertainty about color, not shape, is relevant for color-based categorization[106]) or for different goals (e.g. guide the decision to wager[31] or to update prior estimates[47]), then some aspects of its representation are expected to change. Manipulating the task relevance of uncertainty is thus a promising avenue to explore the function of the representation of uncertainty. In particular, it would be useful to distinguish representations of uncertainty that are automatic and occur independently of task demands from those that are task-dependent[119].

We have stressed that uncertainty can be about different things (e.g. orientation of a grating[35], color[106], the next outcome[103,120–122], probability of an event[47]) and have multiple origins (e.g. prior knowledge, current input). Whether representations of uncertainty are invariant to the origin of this uncertainty, and invariant to what uncertainty is about, remains a largely open question. A related methodological concern, in particular for the code-driven approach, is that $r$ may actually not represent the world state $s$ of interest to the researcher but some other feature $z$; substantial difference between the uncertainty about $s$ and $z$ given $I$ will undermine the code-driven approach. For instance, $r$ in V1 may not represent orientation but instead the intensity of a specific set of image elements present in $I$[82].

As the field matures, a switch from single-model testing to the comparison of different models (e.g. generative models of the observer and the brain used to infer $u(I)$ and $u(r)$; linear vs. non-linear neural codes for $u$ in the correlational approach; code-driven approaches that disentangle the representations of $s$ and $u$) would be valuable to narrow down the neural codes of uncertainty. Because they focus on encoding and decoding respectively, sampling-based codes and probabilistic population codes could also be combined to model processes from input to behavior.

Manipulating prior expectations could help to tackle the pervasive issue of specificity: posterior uncertainty depends on both the current input and the prior, but most studies focus on the former. Manipulating priors enables researchers to partly decorrelate posterior uncertainty from the current input. Some previous studies manipulated priors[50,112] but with the aim of comparing the encoding of the prior and current likelihood. Beyond the methodological interest regarding specificity, systematic manipulation of priors (as in previous behavioral studies[123]) would also be useful in order to study at which stage prior and current uncertainties are combined in the brain when processing the current input, and to compare empirically probabilistic population codes and sampling based codes.

In conclusion, we propose that current studies on the neural representation of uncertainty can be distinguished as code-driven vs. correlational based on whether they rely on assumptions about the neural code of some world state and the accompanying uncertainty. This distinction results in the identification of potentially different types of representation of uncertainty that may be colocalized with, or separated from, the representation of the corresponding world state. Empirical conclusions from both approaches can be assessed with the same set of general criteria, but there is currently an emphasis on different criteria across studies. Since the two approaches differ in the assumptions they require and the type of findings they uncover, there is great potential for them to be used synergistically.

# Acknowledgments

We thank Máté Lengyel and the two other anonymous reviewers for their very helpful comments on the paper. This work was funded by an ANR grant (#18-CE37-0010-01 "CONFI LEARN") and an ERC Starting grant (#947105-NEURAL-PROB) to FM.

# Competing interests

The authors declare no competing interests.

# Data and code availability

No data or code was produced for this review.

# SUPPLEMENTARY INFORMATION

# Studying the neural representations of uncertainty

Edgar Y WALKER[*], Department of Physiology and Biophysics, Computational Neuroscience Center, University of Washington, Seattle, WA

Stephan POHL[*], Department of Philosophy, New York University

Rachel N DENISON, Department of Psychological & Brain Sciences, Boston University, Boston, MA

David L BARACK, Departments of Neuroscience and Philosophy, University of Pennsylvania, Philadelphia, PA

Jennifer LEE, Center for Neural Science, New York University

Ned BLOCK, Department of Philosophy, New York University

Wei Ji MA[#] Center for Neural Science and Department of Psychology, New York University

Florent MEYNIEL[#, §], Cognitive Neuroimaging Unit, INSERM, CEA, CNRS, Université Paris-Saclay, NeuroSpin center, Gif/Yvette, France, florent.meyniel@cea.fr

[*]: These authors contributed equally; [#]: These authors jointly supervised the work; [§]: corresponding author

**Supplementary Table 1: Example studies on the neural representation of uncertainty assessed with general criteria.** Gray: tested and passed. Note that evidence in favor of a criteria is not all-or-none but graded (depending on the strength of effects, sample size, nature of task and measurement etc.); this table is simplified and does not reflect those nuances. The gray color highlights the focus of the studies (unmarked items may still be tested but more secondary). Categorization of studies that include both uncertainty about a latent feature and a future outcome or decision confidence is arbitrary and reflects our chosen emphasis. For the code-driven approach, we distinguish those that assume a probabilistic population code ("PPC") and sampling-based code ("SBC").

**Glossary**

*Uncertainty*: measure of how indeterminate the world state *s* is given the information an observer has about *s*. Under high uncertainty, many different world states are plausible.

*Randomness*: stochasticity in the outcome of a process (synonyms: indeterminacy, noise).

*Decision confidence:* a form of uncertainty about whether a decision is correct or not.

*Neural likelihood function*: the probability of an observed neural activity pattern *r* observed given different hypothesized world states *s*; *s* is the argument of this function.

*Generative model*: a model that specifies how effects follow their causes. It is often characterized only in terms of statistical dependencies, e.g. *p(r|s)*.

*Neural representation of s*: neural activity pattern used by the brain to convey information about *s*.

*Ideal observer*: a model of how the world state *s* is inferred from the input *I* by relying optimally (i.e. with Bayes' rule) on the true generative model of *I*.

| Field | Type | Study | Context | Sensitivity | Specificity | Invariance | Functionality |
|---|---|---|---|---|---|---|---|
| uncertainty about a latent state | code-driven (PPC) | van Bergen et al 2015 | uncertainty about orientation of grating. Human, fMRI | ■ | ■ | | ■ |
| | | van Bergen & Jehee 2019 | uncertainty about orientation of grating. Human, fMRI | | | | ■ |
| | | Geurts et al 2021 | uncertainty about orientation of grating. Human, fMRI | ■ | | | ■ |
| | | Walker et al 2020 | uncertainty about orientation of grating. Monkey, ephys | ■ | ■ | | |
| | | Li et al 2021 | uncertainty about the content of spatial working memory. Human, fMRI | ■ | | | ■ |
| | | Dabney et al 2020 | uncertainty about reward probability. Mice, ephys | ■ | | | ■ |
| | Code-driven (SBC) | Hénaff et al 2020 | uncertainty about a visual feature. Macaque, ephys | ■ | | | |
| | | Orbán et al 2016 | uncertainty about visual stimulus. Macaque and cat ephys | ■ | | ■ | |
| | | Festa et al 2021 | uncertainty about a local feature in an image. Macaque, ephys | ■ | | | |
| | | Echeveste et al 2020 | uncertainty about the orientation of grating. artificial neural networks | ■ | | | |
| | | Haefner et al 2016 | uncertainty about orientation of grating. Artificial neural networks | ■ | | | ■ |
| | | Bányai et al 2019 | uncertainty about local features in an image. Macaque ephys | ■ | | | |
| | correlational | Dekleva et al 2016 | uncertainty about reaching direction. Macaque, ephys | ■ | | | ■ |
| | | Vilares et al 2012 | uncertainty about location of source generating dots. Human, fMRI | ■ | ■ | | ■ |
| | | Bach et al 2011 | uncertainty about proposed gambles. Human, fMRI | ■ | | | ■ |
| | | Meyniel 2020 | uncertainty about generative probability of binary stimulus. Human, MEG | ■ | ■ | | ■ |
| | | Meyniel & Dehaene 2017 | uncertainty about generative probability of binary stimulus. Human, fMRI | ■ | ■ | | ■ |
| | | Payzan-LeNestour et al 2013 | uncertainty about reward probabilities. Human, fMRI | ■ | | | |
| | | O'Reilly et al 2013 | uncertainty about lending location of moving source generating dots. Human, fMRI | ■ | | | ■ |
| | | McGuire et al 2014 | uncertainty about location of source generating dots. Human, fMRI | ■ | | | ■ |
| | | Sedley et al 2016 | uncertainty about mean pitch in sound sequence. Human, ephys | ■ | | | |
| | | Michael et al 2015 | uncertainty about the average stimulus (visual, categorization). Human, fMRI | ■ | | | ■ |
| | | Tomov et al.2020 | uncertainty about reward probabilities. Human, fMRI | ■ | | | |
| | | Badre et al 2012 | uncertainty about reward probability. Human, fMRI | ■ | | | ■ |
| | | Hsu et al 2005 | uncertainty about gambles. Human, fMRI | ■ | | | |
| | | Lebreton et al 2015 | uncertainty about subjective value. Human, fMRI | ■ | | | ■ |
| | | Trudel et al 2021 | uncertainty about the variance of stimuli (visual). Human, fMRI | ■ | | | |
| | | Blankenstein et al 2017 | uncertainty about gambles. Human, fMRI | ■ | | | ■ |
| | | Grinband et al 2006 | uncertainty about categorization boundary (visual). Human, fMRI | ■ | | | |
| | | Tan et al 2016 | uncertainty about the forward sensory-motor model. Human, EEG | ■ | | | |
| | | Huettel 2005 | uncertainty about next observation (shapes). Human, fMRI | ■ | | | |
| | | Ting et al 2015 | uncertainty about the generative proportion of stimuli (visual). Human, fMRI | ■ | | | ■ |
| outcome uncertainty | correlational | FitzGerald et al 2010 | uncertainty about reward delivery (either learned or described). Human, fMRI | | | | ■ |
| | | Strange et al 2005 | uncertainty about next observation (shapes). Human, fMRI | ■ | | | |
| | | Monosov & Hikosaka 2013 | uncertainty about reward delivery. Macaque, ephys | ■ | | ■ | |
| | | Monosov 2017 | uncertainty about reward delivery. Macaque, ephys | ■ | | ■ | |
| | | Monosov et al 2015 | uncertainty about reward delivery or reward size. Macaque, ephys | ■ | | ■ | |
| | | Preuschoff et al 2006 | uncertainty about reward delivery. Human, fMRI | | | | ■ |
| | | Fiorillo 2003 | uncertainty about reward delivery. Macaque, ephys | ■ | | | |
| | | Muller et al 2019 | uncertainty about which option is best (4-arm bandit task). Human, fMRI | ■ | | | ■ |
| | | Nastase et al 2018 | uncertainty about next observation (visual and auditory modalities). Human, fMRI | ■ | | | |
| | | Tzagarakis et al 2010 | uncertainty about which target to reach. MEG human | ■ | | | |
| decision confidence | correlational | Hebart et al 2014 | motion discrimination task. Human, fMRI | ■ | | | ■ |
| | | Gherman & Philiastides 2018 | motion discrimination task. Human, EEG | ■ | | | ■ |
| | | Guggenmos et al 2016 | orientation task. Human, fMRI | ■ | ■ | | ■ |
| | | Kiani & Shadlen 2009 | motion discrimination task. Monkey, ephys | ■ | ■ | ■ | ■ |
| | | Kepecs et al 2008 | odor discrimination task. Rat, ephys | ■ | ■ | ■ | ■ |
| | | Lak et al 2014 | odor discrimination task. Rat, ephys | | ■ | | ■ |
| | | Komura et al 2013 | motion discrimination task. Monkey, ephy | ■ | ■ | | ■ |
| | | Cortese et al 2016 | motion discrimination. Human, fMRI | ■ | ■ | | ■ |
| | | Gherman & Philiastides 2015 | visual object categorization. Human, EEG | ■ | ■ | | |
| | | So & Stuphorn 2016 | confidence about reward delivery. Monkey ephys | ■ | ■ | | ■ |
| | | Odegaard et al 2018 | motion discrimination. Monkey, ephys | ■ | ■ | | ■ |
| | | Masset et al 2020 | Perceptual categorization task (olfactory and auditory stimuli). Rat, ephys | ■ | ■ | ■ | |
| | | Stern et al 2010 | "urns and marbles" task. Human, fMRI | ■ | ■ | | |
| | | de Martino et al 2013 | confidence about utility-based decision. Human, fMRI | ■ | ■ | | |
| | | Bang & Fleming 2018 | motion direction categorization. Human, fMRI | ■ | ■ | | |